\documentclass[francais]{gretsi}
\usepackage[utf8]{inputenc}

\usepackage[T1]{fontenc}

\usepackage{amsmath, amssymb, amsfonts}

\usepackage{amsfonts}
\usepackage{amssymb}
\usepackage{amsthm}
\usepackage{dsfont}
\usepackage{array}
\usepackage{multirow}
\usepackage{lipsum}
\usepackage{mathtools}
\usepackage{cuted}
\usepackage[final]{changes}

\setdeletedmarkup{} 
\definechangesauthor[name={Paul}, color=blue]{pz}
\definechangesauthor[name={Emilie}, color=orange]{ec}
\definechangesauthor[name={Laurent}, color=green]{ld}

\newcommand{\pza}[1]{\added[id=pz]{#1}}
\newcommand{\pzr}[2]{\replaced[id=pz]{#1}{#2}}

\usepackage[caption=false]{subfig}
\usepackage[noadjust]{cite}
\usepackage{bm}
\newcommand{\bs}{\ensuremath{\bm{s}}}
\newcommand{\bpi}{\ensuremath{\bm{\pi}}}
\newcommand{\bt}{\ensuremath{\bm{t}}}
\newcommand{\by}{\ensuremath{\bm{y}}}
\newcommand{\bn}{\ensuremath{\bm{n}}}
\newcommand{\R}{\mathbb{R}}

\newcommand{\pS}{\phantom{*}}

\newtheorem{theorem}{Théorème}

\usepackage{xcolor}

\usepackage{stfloats}
\usepackage{rotating} 

\usepackage{lineno}
\usepackage{siunitx}
\usepackage{etoolbox}
\newcommand{\ubold}{\fontseries{b}\selectfont}
\robustify\ubold

\newcommand{\pend}{PENDANTSS\xspace}

\usepackage[ruled, french,   frenchkw]{algorithm2e}
\begin{document}


\titre{Démélange, déconvolution et débruitage conjoints d'un modèle convolutif parcimonieux avec dérive instrumentale, par pénalisation de  rapports de normes ou quasi-normes lissées (\pend)}

\auteurs{
  \auteur{Paul}{Zheng}{paul.zheng@inda.rwth-aachen.de}{1}
  \auteur{Emilie}{Chouzenoux}{emilie.chouzenoux@centralesupelec.fr}{2}
  \auteur{Laurent}{Duval}{laurent.duval@ifpen.fr}{3}
}

\affils{\affil{1}{Chair   of   Information   Theory   and   Data   Analytics,   RWTH   Aachen   University, Allemagne  }
  \affil{2}{Université Paris-Saclay, CentraleSup\'elec, CVN, Inria, 91190  Gif-sur-Yvette, France   }
  \affil{3}{IFP Energies nouvelles, 92852 Rueil-Malmaison Cedex, France}
}

\resume{Les tâches de débruitage, de filtrage de tendance et de déconvolution sont traditionnellement découplées. Les formulations conjointes se révèlent souvent  des problèmes inverses compliqués, mal posés.  Nous proposons PENDANTSS pour l'isolation conjointe de tendance et la déconvolution, pour des signaux parcimonieux formés de pics. Il fusionne un \emph{a priori} de parcimonie avec l'hypothèse qu'une tendance lisse et le bruit de mesure peuvent être séparés par un filtrage passe-bas. Pour ce faire, nous combinons l'algorithme de séparation de sources ternaire BEADS à des pénalités SOOT/SPOQ en  rapports de normes ou quasi-normes $\ell_p/\ell_q$, promotrices de parcimonie. Un nouvel algorithme efficace est proposé, de type \emph{forward-backward} alterné par blocs à métrique variable avec région de confiance. Il se révèle supérieur à des approches concurrentes pour des mesures classiques appliquées à la chimie analytique. Le code associé est mis à disposition: \url{https://github.com/paulzhengfr/PENDANTSS}.}

\abstract{Denoising, detrending, deconvolution: usual  restoration tasks, traditionally decoupled.  Coupled formulations entail  complex ill-posed inverse problems. We propose \pend for joint trend removal and blind deconvolution of sparse  peak-like signals. It blends a parsimonious  prior with the hypothesis that smooth trend and  noise  can somewhat be separated by low-pass filtering. We  combine the generalized pseudo-norm  ratio SOOT/SPOQ sparse penalties $\ell_p/\ell_q$    with  the BEADS ternary assisted  source separation algorithm. This results in a both   convergent and efficient tool, with  a novel Trust-Region block alternating variable metric forward-backward approach. It outperforms comparable methods, when applied to typically peaked analytical chemistry signals. Reproducible code is provided: \url{https://github.com/paulzhengfr/PENDANTSS}.
}

\maketitle

\section{Contexte}
La base de ce travail prolonge des travaux antérieurs \cite{Duval_L_2015_p-gretsi_suppression_lbdcpapdp,Cherni_A_2019_p-gretsi_forme_lrnlplqspoqrspp}. Récemment publiée dans une revue \cite{Zheng_P_2023_j-ieee-spl_pendantss_pnrdantss}, elle est ici soumise pour la première fois à une conférence. 
Nous considérons le modèle discret de formation de signal suivant :
\begin{equation}
	\bm{y} 
	=   \overline{\bs}\ast \overline{\bpi} + \overline{\bt} + \bn \,.
	\label{eq:observation_signal}
\end{equation}
Il vise à identifier trois composantes : 1) un train parcimonieux d'impulsions  $ \overline{\bs}\in \mathbb{R}^N$, 2) un noyau de convolution en forme de pic $\overline{\bpi}\in \mathbb{R}^L$   et 3) une composante de tendance $\overline{\bt} \in \mathbb{R}^N$ à variations relativement lentes, à partir d'une unique observation $\bm{y} $ bruitée par $\overline{\bn} \in \mathbb{R}^N$.
Ce modèle concerne  une classe courante de données potentiellement multidimensionnelles, dans leur domaine naturel ou après application d'une transformation favorisant la parcimonie \cite{Gauthier_J_2009_j-ieee-tsp_optimization_socfb}. Ce travail se concentre sur le cas de signaux monodimensionnels. 
Il rappelle le problème de soustraction spectrale (en analyse de la parole) visant à séparer des composantes harmoniques (pics dans le domaine de Fourier \cite{Boll_S_1979_j-ieee-tassp_sup_ansss}) d'un fond $ \overline{\bt} + \bn$. Il se retrouve également en analyse biomédicale (ECG, EEG, EMG) ou en astronomie, où les signaux  $\overline{\bm{x}}= \overline{\bs}\ast \overline{\bpi}$ peuvent être nommés lignes ou raies spectrales. La composante $ \overline{\bt}$ concentre des fluctuations  lentes modifiant le niveau de référence (\emph{offset}) des mesures sur  $\overline{\bm{x}}$. Elle peut correspondre à des saisonalités, à des dérives instrumentales liées au  viellisement de capteurs \cite{Barthelme_S_2022_p-gretsi_correction_dimz}, à des variations de calibration. Peu ou mal modélisées (notammment par des modèles paramétriques), un filtrage automatisé de ces tendances, sans altérer les pics, est souvent difficile.
Ce modèle est très courant en analyse physio-chimique (chromatographie, spectrométrie, spectroscopie),  où  $\overline{\bpi}$ prend la forme d'un mélange de  pics positifs à support restreint (gaussiennes, lorentziennes, pseudo-fonction de Voigt). La composante de tendance $\overline{\bt} $ peut s'appeler également  ligne de base, continuum, excursion, fond\ldots

Débruitage et suppression de tendance  conjointes appartiennent à une classe de prétraitements courants de séries temporelles \cite{Baudais_J_2022_p-gretsi_estimation_lbchimlde}, utilisant filtrage, régression paramétrique, remplissage ou désocclusion (\emph{inpainting}). Pour l'analyse de données physico-chimiques, nous renvoyons aux méthodes backcor \cite{Mazet_V_2005_j-chemometr-intell-lab-syst_background_rsdmnqcf} et BEADS \cite{Ning_X_2014_j-chemometr-intell-lab-syst_chromatogram_bedusbeads,Duval_L_2015_p-gretsi_suppression_lbdcpapdp}.
Pour le débruitage et la déconvolution combinées,  mentionnons notamment  \cite{Chaudhuri_S_2014_incoll_blind_dmr,Sun_Q_2021_PREPRINT_convex_sbd} pour des approches promouvant la parcimonie. Nous nous focalisons ici sur les méthodes 
SOOT \cite{Repetti_A_2015_j-ieee-spl_euclid_tsbdsl1l2r} et SPOQ \cite{Cherni_A_2019_p-gretsi_forme_lrnlplqspoqrspp,Cherni_A_2020_j-ieee-tsp_spoq_lpolqrssrams},  employant des rapports de quasi-normes et de normes lissées, présentant une pénalisation avec propriété approchée d'invariance d'échelle.

Afin de résoudre  le problème \eqref{eq:observation_signal}, nous proposons une formulation conjointe non-convexe du problème (section  \ref{Sec_Problem}). Nous présentons un algorithme efficace de séparation basé sur des méthodes \emph{forward-backward}  \cite{Bolte_J_2014_j-math-programm_proximal_almnnp, Chouzenoux_E_2016_j-global-optim_block_cvmfba}, pourvu de preuves de convergence  (section \ref{Sec_Algorithm}). Cet algorithme est évalué --- dans le contexte expérimental décrit en section \ref{Sec_Frame} --- de manière comparative en combinant backcor \cite{Mazet_V_2005_j-chemometr-intell-lab-syst_background_rsdmnqcf} et SOOT/SPOQ \cite{Repetti_A_2015_j-ieee-spl_euclid_tsbdsl1l2r,Cherni_A_2020_j-ieee-tsp_spoq_lpolqrssrams} pour différents niveaux de bruits et de promotion de parcimonie  (section  \ref{Sec_Results}).  

\section{Hypothèses pour la résolution du problème conjoint de démélange\label{Sec_Problem}}
L'équation \eqref{eq:observation_signal} comprend plusieurs inconnues. Tenter de la résoudre impose des hypothèses supplémentaires.
Afin de coupler des différentes  tâches, nous associons à la perte quadratique traditionnelle une régularisation combinée, incorporant certaines hypothèses \emph{a priori}. Le cadre traité par PENDANTSS, pouvant être plus générique, est focalisé ici sur une classe de signaux  observés en analyse physico-chimique. 

Notons  $\iota_A$ la fonction indicatrice de l'ensemble convexe non vide $A$, nulle si son argument appartient à $A$, et identifiée à $+\infty$ sinon. La positivité des pics et du noyau, ainsi que la normalisation de l'intégrale de ce dernier permet de définir dans un premier temps les ensembles $C_1 = [0,+\infty[^N$ et $C_2 \!=\! \mathcal{S}\! = \!\{\bpi \!=\! (\pi_{\ell})_{1 \leq \ell \leq L} \in [0,+\infty[^L \quad \text{t.q.} \quad \sum_{\ell=1}^L \pi_{\ell} = 1  \}$,  limitant (par leurs indicatrices) l'espace de recherche pour le signal parcimonieux et le noyau. 

Nous supposons ensuite que la tendance possède des variations  lentes en regards du bruit. Ainsi par un filtrage passe-bas, il devrait être possible d'en obtenir une estimation correcte. En d'autres termes : en faisant l'hypothèse d'une estimation du signal de pics que l'on puisse soustraire, le bruit résiduel à minimiser par l'attache quadratique aux données  s'exprime par le biais d'un filtre passe-haut $ \bm{H}$:

\begin{equation}
	(\forall \bs \in \mathbb{R}^N) (\forall \bpi \in \mathbb{R}^L)  \;
	\rho(\bs,\bpi)  
 = \frac{1}{2}\|\bm{H}(\by-\bpi * \bs)\|^2.
	\label{eq:rho}
\end{equation}
Cette fonction est Lipschitz différentiable par rapport à $\bs$ (resp. $\bpi$) avec une constante notée $\Lambda_1(\bpi)$ (resp. $\Lambda_2(\bs))$.
La parcimonie du signal  $s$ est favorisée par la  régularisation  $\Psi$  définie (pour $\beta\in]0,+\infty[$)
par la fonction non-convexe :
\begin{equation}
	(\forall \bs \in \mathbb{R}^N) \quad
	\Psi(\bm{s}) = \log\left(\frac{(\ell_{p,\alpha}^p(\bm{s}) + \beta^p)^{1/p}}{\ell_{q,\eta}(\bm{s})} \right),
	\vspace{-.1cm}
\end{equation}
avec les deux approximations paramétriques  de normes ou quasi-normes (de paramètres $(\alpha, \eta)\in]0,+\infty[$) : 
\begin{equation}
	\ell_{p,\alpha}(\bm{s}) = \left( \sum_{n=1}^{N} \left((s_n^2 + \alpha^2)^{p/2}-\alpha^p\right) \right)^{1/p},
	\vspace{-.1cm}
\end{equation}
et
\vspace{-.1cm}
\begin{equation}
	\ell_{q,\eta}(\bm{s}) = \left( \eta^q + \sum_{n=1}^{N} |s_n|^q\right)^{1/q}.
	\vspace{-.1cm}
\end{equation}
Si 
$
	\label{cond:SPOQ}
	q>2, \quad \text{ou} \quad q = 2 \quad \text{et} \quad \eta^2\alpha^{p-2} > \beta^p
$
(ce que nous supposerons), 
alors $\Psi$ est  Lipschitz différentiable sur $\mathbb{R}^N$  et $\bm{0}_N$ (i.e., vecteur de taille $N$ identiquement nul) en est un minimiseur local. Le couple solution $(\widehat{\bpi},\widehat{\bs})$ minimise :
\begin{equation}
	(\forall \bs \in \mathbb{R}^N) (\forall \bpi \in \mathbb{R}^L)  \quad
	\Omega(\bs,\bpi)= f(\bs,\bpi) + g(\bs,\bpi), \label{eq:obj}
\end{equation}
où l'on définit
 \begin{equation}
	(\forall \bs \in \mathbb{R}^N) (\forall \bpi \in \mathbb{R}^L) 
 \begin{cases}
     g(\bs,\bpi)= \iota_{C_1}(\bs) +\iota_{C_2}(\bpi),\\
     f(\bs,\bpi ) = \rho(\bs,\bpi) + \lambda \Psi(\bs).
 \end{cases}
\label{eq:g}
\end{equation}
La tendance est enfin estimée à partir de :
\begin{equation}
 	\widehat{\bt} = (\mathrm{\mathbf{Id}}_N-\bm{H})(\by - \widehat{\bpi}*\widehat{\bs})\,,
	\label{eq:t_hat_orig}
	\vspace{-.15cm}
\end{equation}
avec $\mathrm{\mathbf{Id}}_N$ la matrice identité de $\mathbb{R}^N$.

\section{Algorithme PENDANTSS\label{Sec_Algorithm}}

La structure de  \eqref{eq:obj} suggère l'usage d'une méthode alternée par blocs, mettant à jour séquentiellement la séquence d'impulsions $\bs$ et le noyau $\bpi$. PENDANTSS s'appuie sur l'algorithme TR-BC-VMFB (Alg.~\ref{algo:TR-BC-VMFB}), qui généralise le BC-VMFB  \cite{Chouzenoux_E_2016_j-global-optim_block_cvmfba}  employé dans  \cite{Repetti_A_2015_j-ieee-spl_euclid_tsbdsl1l2r} en déconvolution aveugle.

Soient $(\bs_k,\bpi_k)$ les estimées de l'itération $k \in \mathbb{N}$. Le calcul de $\bs_{k+1}$ s'obtient par une étape de VMFB \cite{Chouzenoux_E_2014_j-optim-theory-appl_variable_mfbamsdfcf}, accélérée par un schéma de région de confiance. 
Nous introduisons tout d'abord la métrique MM (de majoration-minimisation) :
\begin{multline}
	\bm{A}_{1,\rho}(\bm{s}_k,\bpi_k) = (\Lambda_1(\bpi_k) + \lambda \chi_{q,\rho}) \mathrm{\mathbf{Id}}_N + \\ \frac{\lambda}{\ell_{p,\alpha}^p(\bs_k) + \beta^p}\text{Diag}((s_{n,k}^2+\alpha^2)^{p/2-1})_{1\leq n\leq N},
	\label{eq:A_qrho}
\end{multline}
avec la constante  $
	\chi_{q,\rho} = \frac{q-1}{(\eta^q+\rho^q)^{2/q}}
$. On construit  une majoration de  \eqref{eq:obj} par rapport 
à la variable $\bs$ (voir \cite[Prop. 2]{Cherni_A_2020_j-ieee-tsp_spoq_lpolqrssrams}) :
\begin{multline}
	\label{eq:maj1}
	(\forall \bs \in \bar{\mathcal{B}}_{q,\rho} \cap C_1) \quad
	\Omega(\bs,\bpi_k) \leq f(\bs_k,\bpi_k) \\+ {(\bs - \bs_k)}^\top \nabla_1 f(\bs_k,\bpi_k)  + \frac{1}{2} \| \bs - \bs_k\|^2_{\bm{A}_{1,\rho}(\bs_k,\bpi_k)},
\end{multline}
avec, pour tout $\bm{z}\in \R^N$, $\|\bm{z}\|_{\bm{A}} = (\bm{z}^\top \bm{A}\bm{z})^{1/2}$. Le domaine de validité de \eqref{eq:maj1} est limité par le complément de la boule $\ell_q$, 
\begin{equation}	\bar{\mathcal{B}}_{q,\rho}   =\{\bm{s}=(s_n)_{1\leq n\leq N} \in\R^N | \sum_{n=1}^{N} |s_n|^q \geq \rho^q \}.
\label{eq:TR_ballex}
\end{equation}
 
Nous introduisons donc un schéma de région de confiance  (\emph{Trust-Region} ou TR \cite{Conn_A_2000_book_trust-region_m}), permettant de contrôler le domaine des itérés.
Soit $\mathcal{I} >0$, un nombre maximal de tests de régions de confiance, et $(\rho_{k,i})_{1 \leq i \leq \mathcal{I}}$ une liste de rayons testés :  
\begin{equation}
	\vspace{-0.2cm}
	\rho_{k,i} = \begin{cases}
		\sum_{n=1}^{N}|s_{n,k}|^q & \text{si } i = 1\,,\\
		\theta \rho_{k,i-1} & \text{si } 2\leq i\leq \mathcal{I}-1\,,\\
		0 & \text{si } i = \mathcal{I}\,.
	\end{cases}
	\label{eq:TRradius}
\end{equation}
On calcule alors la matrice MM associée $\bm{A}_{1,\rho_{k,i}}(\bm{s}_k,\bpi_k)$, et l'on définit $\bm{s}_{k,i}$ comme minimiseur du terme de droite de~\eqref{eq:maj1}. La boucle TR s'interrompt dès que $\bm{s}_{k,i} \in  \bar{\mathcal{B}}_{q,\rho_{k,i}}$, et définit ainsi $\bs_{k+1}$. En général, la minimisation de la majorante \eqref{eq:maj1} n'admet pas de solution explicite. Néanmoins, par notre choix de $ C_1$ et la structure diagonale de \eqref{eq:A_qrho}, la résolution  est directe, d'après \cite[Prop. 24.11]{Bauschke_H_2011_book_convex_amoths} et \cite[Cor. 9]{Becker_S_2012_p-nips_quasi-newton_psm},
$(\forall i \in \{1,\ldots,\mathcal{I}\})$
\begin{equation}
\label{eq:updateS}
	\bs_{k,i} \!=\! \text{Proj}_{ C_1}\!\!\left( \bs_k \!-\! \gamma_{s,k}\bm{A}_{1,\rho_{k,i}}(\bm{s}_{k},\bm{\pi}_{k})^{-1}\nabla_1 f(\bs_k,\bpi_k) \right).
	\end{equation}

La mise à jour du noyau  s'exprime simplement, par une étape de descente de gradient projeté : 
\begin{equation}
\bpi_{k+1} = \text{Proj}_{\mathcal{S}}\left(
\bpi_k - \gamma_{\pi,k}\Lambda_{2}(\bm{s}_{k+1})^{-1}\nabla_2 f(\bs_{k+1},\bpi_k)
\right),
\label{eq:updatePi}
\end{equation}
avec ($\text{Proj}_{\mathcal{S}}$) la projection sur le simplexe unité, pour laquelle il existe des méthodes de calcul rapides~\cite{Condat_L_2016_j-math-programm_fast_psl1b}. La méthode (incluant les domaines de validité des pas $(\gamma_{s,k},\gamma_{\pi,k})_{k \in \mathbb{N}}$), est résumée dans l'algorithme \ref{algo:TR-BC-VMFB}, dont la convergence est assurée.
\begin{theorem}
	\label{prop:conv_BD}
	Soient $(\bm{s}_k,\bm{\pi}_k)_{k \in \mathbb{N}}$ générés par l'algorithme~\ref{algo:TR-BC-VMFB}. Si $C_1$ et  $C_2$ sont des ensembles semi-algébriques (ce qui est vérifié dans notre cas), alors la suite  $(\bm{s}_{k}, \bm{\pi}_k)_{k \in \mathbb{N}}$ converge  vers un point critique $(\widehat{\bm{s}}, \widehat{\bm{\pi}})$ de \eqref{eq:obj}. 
\end{theorem}

\begin{algorithm}
	\SetAlgoLined
	\textbf{Paramètres :} $K_{\max}>0$, $\varepsilon>0$, $\mathcal{I} > 0$,~$\theta\in]0,1[$, pas  $(\gamma_{s,k})_{k \in \mathbb{N}} \in [\underline{\gamma}, 2-\overline{\gamma}]$ et  $(\gamma_{\pi,k})_{k \in \mathbb{N}}  \in [\underline{\gamma}, 2- \overline{\gamma}]$ pour un choix admissible de  $(\underline{\gamma}, \overline{\gamma}) \in ]0,+\infty[^2$, $(p,q) \in ]0,2[\times [2,+\infty[$,\\
	ensembles convexes $(C_1,C_2) \subset \mathbb{R}^N \times \mathbb{R}^L$.\\
	\textbf{initialisation:}~$\bm{s}_0\in C_1$,~$\bm{\pi}_0\in C_2$\\
	
	\Pour{$k = 0, 1,\ldots$}{
		\underline{\emph{mise à jour du signal}}\\
		\Pour{$i = 1,\ldots, \mathcal{I}$}{
			Définition du rayon $\rho_{k,i}$, d'après \eqref{eq:TRradius} \;
			Définition de la métrique MM avec~\eqref{eq:A_qrho} et calcul de $\bs_{k,i}$ par \eqref{eq:updateS}.\\ 
			\Si{$\bm{s}_{k,i}\in \bar{\mathcal{B}}_{q,\rho_{k,i}}$}{sortie de boucle}
		}
		$\bm{s}_{k+1} = \bm{s}_{k,i}$\;
		\underline{\emph{mise à jour du noyau}}\\
  Calcul de $\bpi_{k+1}$ par \eqref{eq:updatePi}.\\
		\underline{\emph{critère d'arrêt}}\\
		
		\Si{$\|\bs_k -\bs_{k+1}|| \leq \varepsilon$ \text{ou} $k \geq K_{\max}$}{sortie de boucle}
	}	
	$(\widehat{\bs},\widehat{\bpi}) = (\bs_{k+1},\bpi_{k+1})$ 
	et 
	$\bm{\widehat{t}}$
	 selon \eqref{eq:t_hat_orig}
	
	\Sortie{$\widehat{\bs},\widehat{\bpi}, \bm{\widehat{t}}$}
	\caption{TR-BC-VMFB pour minimiser \eqref{eq:obj}.}
	\label{algo:TR-BC-VMFB}
\end{algorithm}

\section{Contexte de validation expérimentale\label{Sec_Frame}}

\begin{figure}[t]
	\centering
	\subfloat[Jeu de données C.\label{fig:dataA}]{
		\includegraphics[width=0.75\linewidth]{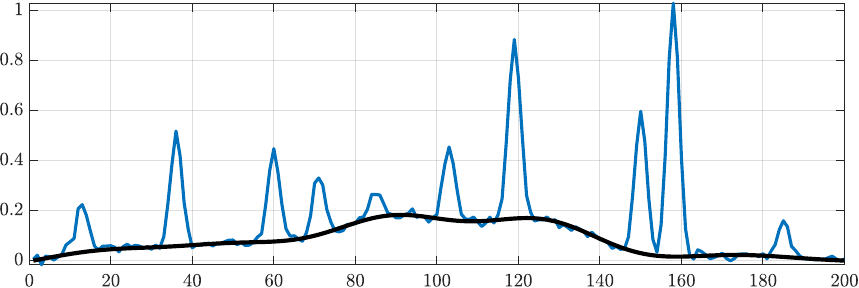}}
	\vspace{-.2cm}
	\hfill 
	\subfloat[Signal impulsif parcimonieux (jeu de données  C).\label{fig:dataA}]{
		\includegraphics[width=0.75\linewidth]{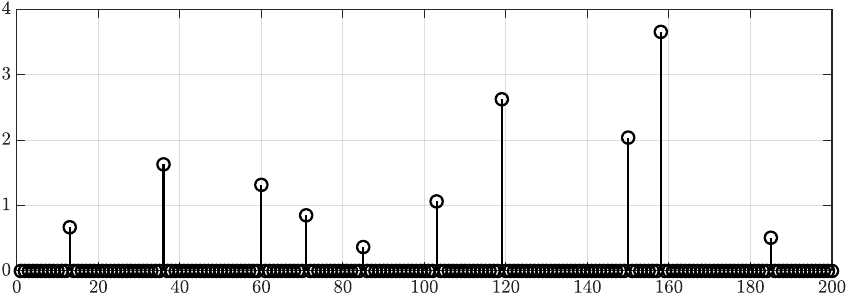}}
	\vspace{-.2cm}
	\hfill 
	\subfloat[Jeu de données D.\label{fig:dataB}]{
		\includegraphics[width = 0.75\linewidth]{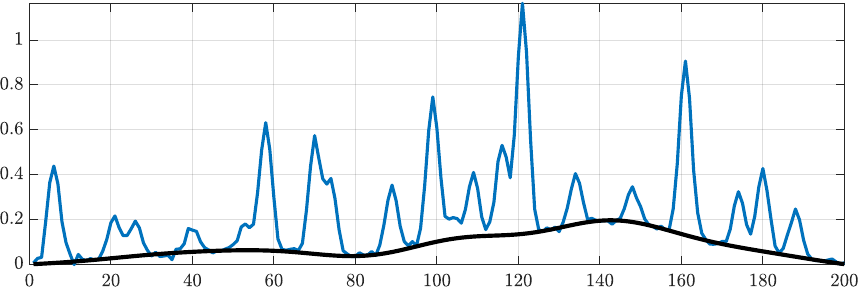}}
	\vspace{-.2cm}
	\hfill 
	\subfloat[Signal impulsif parcimonieux (jeu de données  D).\label{fig:dataB}]{
		\includegraphics[width = 0.75\linewidth]{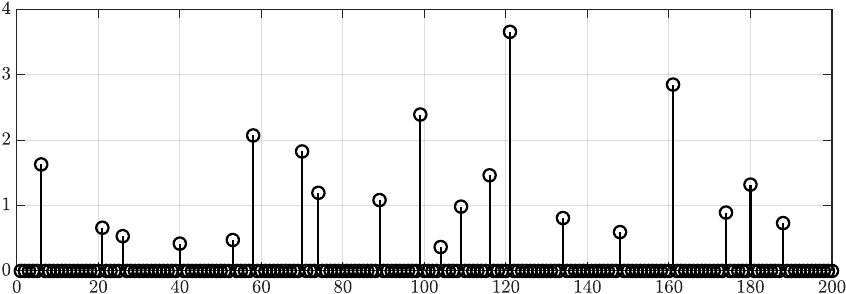}}
	\vspace{-.1cm}
	\caption{Signal observé $\bm{y}$ (bleu) et tendance/ligne de base $\overline{\bt}$ (noir): (a) et (c). Signal parcimonieux original  inconnu~$\overline{\bs}$ : (b) et (d). Les signaux C et D comptent 10 et 20 impulsions respectivement (\SI{5}{\percent} et   \SI{10}{\percent} de parcimonie).}
	\label{Fig_Dataset}
	\vspace{-.2cm}
\end{figure}

Nous considérons les jeux de données nommés C et D. Les signaux parcimonieux originaux $\overline{\bs}$ et les observations $\by$ sont représentés dans la figure \ref{Fig_Dataset},de longueur $N = 200$. Les  observations $\by$ sont obtenues à partir de \eqref{eq:observation_signal}, avec un noyau $\overline{\bpi}$ défini par une fonction gaussienne normalisée, d'écart-type $0,15$ et de support tronqué de  taille $L = 21$. Le bruit est blanc, gaussien,  à moyenne nulle. Son niveau $\sigma$ est fixé à un pourcentage variable de l'amplitude maximale  $x_{\max}$  de $\overline{\bm{x}} = \overline{\bpi} \ast \overline{\bs}$, convolution implémentée par bourrage de zéros.

Les paramètres de l'algorithme PENDANTSS ont été choisis constants par soucis de simplicité :  $ \gamma_{s,k} \equiv 1.9$ et  $\gamma_{\pi,k} \equiv 1.9$ satisfont la contrainte d'intervalle. Nous avons choisi $\theta = 0.5$ comme incrément du rayon de confiance, ainsi qu'un maximum de $\mathcal{I} = 50$ tests. L'initialisation pour toutes les méthodes suit celle proposée dans \cite{Repetti_A_2015_j-ieee-spl_euclid_tsbdsl1l2r} : $\bs_0 \in C_1$ est un signal constant, positif, $\bpi_0 \in C_2$ est un filtre gaussien de largeur 1. Les critères d'arrêt sont : $\varepsilon = 10^{-6}\sqrt{N}$ et $K_{\max} = 3000$.

Les hyperparamètres --- régularisation pour backcor~\cite{Mazet_V_2005_j-chemometr-intell-lab-syst_background_rsdmnqcf} et ceux de  SPOQ/SOOT  $(\lambda, \alpha,\beta,\eta)$ --- ont été ajustés à partir d'une seule réalisation de référence, en employant une métrique composite, combinant les trois composantes-cibles (train d'impulsions, noyau, tendance) : $2\text{SNR}_{\bs} + \text{SNR}_{\bpi} + \text{SNR}_{\bt}$. \pzr{Suivant le même critère composite, la}{La} fréquence de coupure \pza{$f_c$} du filtre passe-haut, dans \eqref{eq:t_hat_orig}, résulte du choix du meilleur point, parmi les dix \pzr{premiers pics}{premières valeurs} du module du spectre de fréquence du signal. \pza{Parmi les hyperparamètres, $\alpha$ peut aisément être choisi constant (typiquement~$\alpha=7\times 10^{-7}$ pour nos données).}

Du fait de l'ambigüité de position classique en démélange, pour s'assurer que le noyau est correctement centré, nous appliquons au noyau estimé  un post-traitement du décalage spatial entier, pour nous assurer qu'il est correctement centré.  Une recherche par quadrillage (\emph{grid search}) détermine le nombre de boucles permettant de maximiser le $\text{SNR}_{\bs}$ du train d'impulsion parcimonieux. 

\vspace{-.2cm}
\section{Résultats numériques\label{Sec_Results}}
\vspace{-.1cm}
Nous comparons les performances de PENDANTSS par ablation vis-à-vis  (i) de la référence backcor~\cite{Mazet_V_2005_j-chemometr-intell-lab-syst_background_rsdmnqcf} en suppression de tendance\pza{ avec une recherche par grille ou quadrillage pour optimiser l'ordre du polynôme et le seuil}, suivie de la déconvolution  aveugle proposée dans \cite{Repetti_A_2015_j-ieee-spl_euclid_tsbdsl1l2r} pour estimer le signal $\widehat{\bs}$ et le noyau 
$\widehat{\bpi}$, (ii) du flot complet de PENDANTSS avec les paramètres SPOQ $(p,q) = (1,2)$ (soit SOOT) ou $(p,q) = (0.75,10)$. 
Les différentes estimations  sont évaluées en rapport signal-sur-bruit (SNR) : signal parcimonieux (SNR$_{\bs}$), noyau (SNR$_{\bpi}$) et tendance (SNR$_{\bt}$). En particulier, SNR$_{\bs} = 20\log_{10}(\|\overline{\bs}\|_2 / \|\overline{\bs} - \widehat{\bs}\|_2)$. Nous évaluons de surcroît le $\text{TSNR}$, correspondant à la même quantité, évaluée uniquement sur le support (supposé connu) du signal parcimonieux original. Ce support n'est pas connu en général, cependant cette mesure est utile pour quantifier la performance de tâches d'estimation de quantités ancillaires, calculées sur les pics (amplitude, largeur, surface). De telles mesures, importantes en analyse physico-chimique quantitative, sont sensibles au filtrage de tendance et à la déconvolution. 

Les résultats sont résumés dans la table \ref{tab: results}. Nous reportons les quantités moyennes, et leur écart-type (après le signe ``$\pm$"), sur 30 réalisations.
Le meilleur résultat et le second sont identifiés par deux (**) ou une  (*) étoiles. Dans la plupart des situations, PENDANTSS se révèle supérieur aux approches découplées. Il convient néanmoins de rester mesuré en regard des écarts-types parfois importants, ce qui n'est pas surprenant dans l'évaluation de métriques quadratiques pour des signaux de nature parcimonieuse. 

\vspace{-.2cm}
\section{Conclusion et perspectives}
\vspace{-.1cm}
Nous proposons l'algorithme PENDANTSS, pour résoudre le problème compliqué de séparation de tendance, conjoint à une déconvolution aveugle parcimonieuse. La méthode prend en compte une hypothèse de dérive lisse "basse fréquence" en l'incorporant à un problème de déconvolution aveugle. Ce dernier s'appuie sur l'usage récent de pénalité en rapport de quasi-normes ou normes de type SOOT/SPOQ. La validation proposée indique un gain quantitatif par rapport aux méthodes de référence, pour des signaux parcimonieux positifs, comme l'on en rencontre en chimie analytique. 

Il reste à étendre la validation à de plus larges classes de signaux parcimonieux, ainsi qu'à proposer des estimations plus intuitives des hyper-paramètres, en fonction de la nature des signaux parcimonieux analysés, et particulièrement en regard de critères de séparabilité des signaux de pics.
 Un code en langage Matlab est mis à disposition \url{https://github.com/paulzhengfr/PENDANTSS}.

\section*{Remerciement}
Ce travail a bénéficié de la bourse ERC  (\emph{European Research Council, Starting Grant}) MAJORIS ERC-2019-STG-850925.

\begin{table}[t]
	\tiny
	\renewcommand{\arraystretch}{1.3}
	{
		\centering
		\caption{Résultats numériques pour les données C et D. Quantités exprimées en rapport signal-sur-bruit (SNR) en décibels (dB). La méthode la plus efficace est suivie de~**, la seconde par~*. \label{tab: results}}
		\centering
		\begin{tabular}{|c| r  | r | r  | r| r | }
			\cline{3-6}
			\multicolumn{2}{c}{}& \multicolumn{2}{|c|}{Donnée C}   & \multicolumn{2}{|c|}{Donnée D}  \\ \hline 
			\multicolumn{2}{|c|}{Niveau de bruit $\sigma$ (\SI{}{\percent} de $x_{\max})$}& \SI{0.5}{\percent}\pS\pS & \SI{1}{\percent}\pS\pS   & \SI{0.5}{\percent}\pS\pS & \SI{1}{\percent}\pS\pS  \\ \hline\hline
			\parbox[t]{2mm}{\multirow{4}{*}{\rotatebox[origin=c]{90}{SNR$_{\bs}$}}}

			& backcor+SOOT   & \num{33.224638}$\pm$\num{4.457540}\pS\pS &  \num{26.901807}$\pm$\num{2.957293}\pS\pS & \num{28.276753}$\pm$\num{1.060368}\pS\pS & \num{21.598744}$\pm$\num{2.176613}\pS\pS
			
			\\\cline{2-6}
			& backcor+SPOQ   & \num{31.979984}$\pm$\num{5.216894}\pS\pS &  \num{26.408462}$\pm$\num{3.099133}\pS\pS & \num{25.502839}$\pm$\num{3.906442}\pS\pS & \num{19.402096}$\pm$\num{4.066848}\pS\pS

			\\\cline{2-6}
			& PENDANTSS  (1, 2)   & \num{34.898926}$\pm$\num{2.048982}*\pS &  \num{29.640298}$\pm$\num{1.976187}** & \num{35.080852}$\pm$\num{1.665643}** &  \num{30.130005}$\pm$\num{1.961174}**

			\\\cline{2-6}
			& PENDANTSS  (0.75, 10)    &      \num{35.238868}$\pm$\num{5.702458}** &  \num{29.211855}$\pm$\num{6.215003}*\pS & \num{34.680017}$\pm$\num{5.231616}*\pS &  \num{28.309922}$\pm$\num{4.353153}*\pS 
			

			\\
			\hline \hline
			
			\parbox[t]{2mm}{\multirow{4}{*}{\rotatebox[origin=c]{90}{TSNR$_{\bs}$}}}
			& backcor+SOOT  & \num{33.224638}$\pm$\num{4.457540}\pS\pS &  \num{26.901807}$\pm$\num{2.957293}\pS\pS & \num{31.202335}$\pm$\num{1.356333}\pS\pS & \num{26.535885}$\pm$\num{2.917023}\pS\pS  
			
			\\\cline{2-6}
			& backcor+SPOQ   & \num{31.979984}$\pm$\num{5.216894}\pS\pS &  \num{26.408462}$\pm$\num{3.099133}\pS\pS & \num{26.043288}$\pm$\num{3.578244}\pS\pS & \num{22.145073}$\pm$\num{3.857987}\pS\pS
			\\\cline{2-6}
			
			& PENDANTSS  (1, 2)   &  \num{36.411096}$\pm$\num{2.365084}\pS\pS &  \num{32.243503}$\pm$\num{2.285112}** & \num{35.639936}$\pm$\num{1.440617}** & \num{31.024606}$\pm$\num{1.864290}**

			\\\cline{2-6}
			& PENDANTSS  (0.75, 10)    &\num{36.558058}$\pm$\num{5.357315}** &  \num{30.235316}$\pm$\num{5.477227}*\pS & \num{35.249299}$\pm$\num{4.609214}*\pS & \num{28.654551}$\pm$\num{3.623507}*\pS

			\\
			\hline \hline

			\parbox[t]{2mm}{\multirow{4}{*}{\rotatebox[origin=c]{90}{SNR$_{\bt}$}}}
			& backcor+SOOT   & \num{24.589672}$\pm$\num{1.578547}\pS\pS &  \num{20.593675}$\pm$\num{1.037798}\pS\pS & \num{20.168002}$\pm$\num{0.987961}\pS\pS & \num{17.237666}$\pm$\num{1.406592}\pS\pS  
			
			\\\cline{2-6}
			& backcor+SPOQ   & \num{24.589672}$\pm$\num{1.578547}\pS\pS &  \num{20.593675}$\pm$\num{1.037798}\pS\pS & \num{20.168002}$\pm$\num{0.987961}\pS\pS & \num{17.237666}$\pm$\num{1.406592}\pS\pS  
			
			\\\cline{2-6}
			
			& PENDANTSS  (1, 2)   & \num{30.871136}$\pm$\num{1.321580}*\pS &  \num{26.035330}$\pm$\num{1.260890}*\pS & \num{27.407441}$\pm$\num{0.977796}*\pS & \num{25.005354}$\pm$\num{1.648530}**

			\\\cline{2-6}
			&PENDANTSS  (0.75, 10)     &\num{31.103622}$\pm$\num{1.760293}** &  \num{26.241737}$\pm$\num{1.118892}** & \num{27.830191}$\pm$\num{1.164385}** & \num{22.433647}$\pm$\num{1.420979}*\pS

			\\
			\hline \hline
			
			\parbox[t]{2mm}{\multirow{4}{*}{\rotatebox[origin=c]{90}{SNR$_{\bpi}$}}}
			& backcor+SOOT   &\num{42.103715}$\pm$\num{2.845398}\pS\pS  &  \num{33.558193}$\pm$\num{2.286156}\pS\pS  & \num{44.504748}$\pm$\num{2.547148}*\pS & \num{37.839715}$\pm$\num{2.856357}\pS\pS   
			
			\\\cline{2-6}
			& backcor+SPOQ   & \num{42.105335}$\pm$\num{2.857194}\pS\pS  &  \num{33.140738}$\pm$\num{2.255803}\pS\pS  & \num{40.527403}$\pm$\num{2.996164}\pS\pS  & \num{39.382836}$\pm$\num{2.098684}**  
			
			\\\cline{2-6}
			
			& PENDANTSS  (1, 2)  & \num{42.593702}$\pm$\num{3.088251}** &  \num{37.379322}$\pm$\num{2.237350}** & \num{44.226005}$\pm$\num{2.553586}*\pS & \num{38.901575}$\pm$\num{2.548714}\pS\pS

			\\\cline{2-6}
			& PENDANTSS  (0.75, 10)     &
			\num{42.329755}$\pm$\num{2.421032}*\pS &  \num{36.998031}$\pm$\num{2.088651}*\pS & \num{44.957264}$\pm$\num{2.629747}** & \num{38.637602}$\pm$\num{2.592465}\pS\pS

			\\
			\hline
		\end{tabular}
	}
\end{table}




\small
\begin{thebibliography}{10}
	\expandafter\ifx\csname fonteauteurs\endcsname\relax
	\def\fonteauteurs{\scshape}\fi
	
	\bibitem{Barthelme_S_2022_p-gretsi_correction_dimz}
	S.~\bgroup\fonteauteurs\bgroup Barthelme\egroup\egroup{},
	F.~\bgroup\fonteauteurs\bgroup Chatelain\egroup\egroup{},
	C.~\bgroup\fonteauteurs\bgroup Cascales\egroup\egroup{} et
	C.~\bgroup\fonteauteurs\bgroup Herrier\egroup\egroup{} :
	\newblock Correction de dérive pour l'interférométrie de {M}ach-{Z}ehnder.
	\newblock \emph{In} {\em Proc. GRETSI}, 2022.
	
	\bibitem{Baudais_J_2022_p-gretsi_estimation_lbchimlde}
	J.-Y. \bgroup\fonteauteurs\bgroup Baudais\egroup\egroup{},
	M.~\bgroup\fonteauteurs\bgroup Leclerc\egroup\egroup{} et
	C.~\bgroup\fonteauteurs\bgroup Langrume\egroup\egroup{} :
	\newblock Estimation de ligne de base de capteurs d’humectation :
	intégration et minimums locaux à différentes échelles.
	\newblock \emph{In} {\em Proc. GRETSI}, 2022.
	
	\bibitem{Bauschke_H_2011_book_convex_amoths}
	H.~H. \bgroup\fonteauteurs\bgroup Bauschke\egroup\egroup{} et P.~L.
	\bgroup\fonteauteurs\bgroup Combettes\egroup\egroup{} :
	\newblock {\em Convex analysis and monotone operator theory in {H}ilbert
		spaces}.
	\newblock CMS books in mathematics. Springer, 2e \'edition, 2011.
	
	\bibitem{Becker_S_2012_p-nips_quasi-newton_psm}
	S.~\bgroup\fonteauteurs\bgroup Becker\egroup\egroup{} et M.~J.
	\bgroup\fonteauteurs\bgroup Fadili\egroup\egroup{} :
	\newblock A quasi-{N}ewton proximal splitting method.
	\newblock \emph{In} {\em Proc. Ann. Conf. Neur. Inform. Proc. Syst.}, volume~2,
	pages 2618--2626, Dec. 3-6, 2012.
	
	\bibitem{Boll_S_1979_j-ieee-tassp_sup_ansss}
	S.~\bgroup\fonteauteurs\bgroup Boll\egroup\egroup{} :
	\newblock Suppression of acoustic noise in speech using spectral subtraction.
	\newblock {\em IEEE Trans. Acoust. Speech Signal Process.},
	27(2)\string:\penalty500\relax 113--120, Apr. 1979.
	
	\bibitem{Bolte_J_2014_j-math-programm_proximal_almnnp}
	J.~\bgroup\fonteauteurs\bgroup Bolte\egroup\egroup{},
	S.~\bgroup\fonteauteurs\bgroup Sabach\egroup\egroup{} et
	M.~\bgroup\fonteauteurs\bgroup Teboulle\egroup\egroup{} :
	\newblock Proximal alternating linearized minimization for nonconvex and
	nonsmooth problems.
	\newblock {\em Math. Programm.}, 146(1-2)\string:\penalty500\relax 459--494,
	Aug. 2014.
	
	\bibitem{Chaudhuri_S_2014_incoll_blind_dmr}
	S.~\bgroup\fonteauteurs\bgroup Chaudhuri\egroup\egroup{},
	R.~\bgroup\fonteauteurs\bgroup Velmurugan\egroup\egroup{} et
	R.~\bgroup\fonteauteurs\bgroup Rameshan\egroup\egroup{} :
	\newblock Blind deconvolution methods: A review.
	\newblock \emph{In} {\em Blind Image Deconvolution. Methods and Convergence},
	pages 37--60. Springer, 2014.
	
	\bibitem{Cherni_A_2020_j-ieee-tsp_spoq_lpolqrssrams}
	A.~\bgroup\fonteauteurs\bgroup Cherni\egroup\egroup{},
	E.~\bgroup\fonteauteurs\bgroup Chouzenoux\egroup\egroup{},
	L.~\bgroup\fonteauteurs\bgroup Duval\egroup\egroup{} et J.-C.
	\bgroup\fonteauteurs\bgroup Pesquet\egroup\egroup{} :
	\newblock {SPOQ} $\ell_p$-over-$\ell_q$ regularization for sparse signal
	recovery applied to mass spectrometry.
	\newblock {\em IEEE Trans. Signal Process.}, 68\string:\penalty500\relax
	6070--6084, 2020.
	
	\bibitem{Cherni_A_2019_p-gretsi_forme_lrnlplqspoqrspp}
	A.~\bgroup\fonteauteurs\bgroup Cherni\egroup\egroup{}, E.
	\bgroup\fonteauteurs\bgroup Chouzenoux\egroup\egroup{},
	L.~\bgroup\fonteauteurs\bgroup Duval\egroup\egroup{} et J.-C.
	\bgroup\fonteauteurs\bgroup Pesquet\egroup\egroup{} :
	\newblock Forme lissée de rapports de normes $\ell_p$/$\ell_q$ ({SPOQ}) pour
	la reconstruction des signaux avec pénalisation parcimonieuse.
	\newblock \emph{In} {\em Proc. GRETSI}, 2019.
	
	\bibitem{Chouzenoux_E_2014_j-optim-theory-appl_variable_mfbamsdfcf}
	E.~\bgroup\fonteauteurs\bgroup Chouzenoux\egroup\egroup{}, J.-C.
	\bgroup\fonteauteurs\bgroup Pesquet\egroup\egroup{} et
	A.~\bgroup\fonteauteurs\bgroup Repetti\egroup\egroup{} :
	\newblock Variable metric forward-backward algorithm for minimizing the sum of
	a differentiable function and a convex function.
	\newblock {\em J. Optim. Theory Appl.}, 162(1)\string:\penalty500\relax
	107--132, Jul. 2014.
	
	\bibitem{Chouzenoux_E_2016_j-global-optim_block_cvmfba}
	E.~\bgroup\fonteauteurs\bgroup Chouzenoux\egroup\egroup{}, J.-C.
	\bgroup\fonteauteurs\bgroup Pesquet\egroup\egroup{} et A.~
	\bgroup\fonteauteurs\bgroup Repetti\egroup\egroup{} :
	\newblock A block coordinate variable metric forward-backward algorithm.
	\newblock {\em J. Global Optim.}, 66(3)\string:\penalty500\relax 457--485, Feb.
	2016.
	
	\bibitem{Condat_L_2016_j-math-programm_fast_psl1b}
	L.~\bgroup\fonteauteurs\bgroup Condat\egroup\egroup{} :
	\newblock Fast projection onto the simplex and the $l_1$ ball.
	\newblock {\em Math. Programm.}, 158(1-2)\string:\penalty500\relax 575--585,
	2016.
	
	\bibitem{Conn_A_2000_book_trust-region_m}
	A.~R. \bgroup\fonteauteurs\bgroup Conn\egroup\egroup{}, N.~I.~M.
	\bgroup\fonteauteurs\bgroup Gould\egroup\egroup{} et P.~L.
	\bgroup\fonteauteurs\bgroup Toint\egroup\egroup{} :
	\newblock {\em Trust-Region Methods}.
	\newblock MOS-SIAM Series on Optimization. Society for Industrial Mathematics,
	2000.
	
	\bibitem{Duval_L_2015_p-gretsi_suppression_lbdcpapdp}
	L.~\bgroup\fonteauteurs\bgroup Duval\egroup\egroup{},
	A.~\bgroup\fonteauteurs\bgroup Pirayre\egroup\egroup{},
	X.~\bgroup\fonteauteurs\bgroup Ning\egroup\egroup{} et
	I.~\bgroup\fonteauteurs\bgroup Selesnick\egroup\egroup{} :
	\newblock Suppression de ligne de base et débruitage de chromatogrammes par
	pénalisation asymétrique de positivité et dérivées parcimonieuses.
	\newblock \emph{In} {\em Proc. GRETSI}, 2015.
	
	\bibitem{Gauthier_J_2009_j-ieee-tsp_optimization_socfb}
	J.~\bgroup\fonteauteurs\bgroup Gauthier\egroup\egroup{},
	L.~\bgroup\fonteauteurs\bgroup Duval\egroup\egroup{} et J.-C.
	\bgroup\fonteauteurs\bgroup Pesquet\egroup\egroup{} :
	\newblock Optimization of synthesis oversampled complex filter banks.
	\newblock {\em IEEE Trans. Signal Process.}, 57(10)\string:\penalty500\relax
	3827--3843, Oct. 2009.
	
	\bibitem{Mazet_V_2005_j-chemometr-intell-lab-syst_background_rsdmnqcf}
	V.~\bgroup\fonteauteurs\bgroup Mazet\egroup\egroup{},
	C.~\bgroup\fonteauteurs\bgroup Carteret\egroup\egroup{},
	D.~\bgroup\fonteauteurs\bgroup Brie\egroup\egroup{},
	J.~\bgroup\fonteauteurs\bgroup Idier\egroup\egroup{} et
	B.~\bgroup\fonteauteurs\bgroup Humbert\egroup\egroup{} :
	\newblock Background removal from spectra by designing and minimising a
	non-quadratic cost function.
	\newblock {\em Chemometr. Intell. Lab. Syst.}, 76(2)\string:\penalty500\relax
	121--133, 2005.
	
	\bibitem{Ning_X_2014_j-chemometr-intell-lab-syst_chromatogram_bedusbeads}
	X.~\bgroup\fonteauteurs\bgroup Ning\egroup\egroup{}, I.~W.
	\bgroup\fonteauteurs\bgroup Selesnick\egroup\egroup{} et
	L.~\bgroup\fonteauteurs\bgroup Duval\egroup\egroup{} :
	\newblock Chromatogram baseline estimation and denoising using sparsity
	({BEADS}).
	\newblock {\em Chemometr. Intell. Lab. Syst.}, 139\string:\penalty500\relax
	156--167, Dec. 2014.
	
	\bibitem{Repetti_A_2015_j-ieee-spl_euclid_tsbdsl1l2r}
	A.~\bgroup\fonteauteurs\bgroup Repetti\egroup\egroup{}, M.~Q.
	\bgroup\fonteauteurs\bgroup Pham\egroup\egroup{},
	L.~\bgroup\fonteauteurs\bgroup Duval\egroup\egroup{},
	E.~\bgroup\fonteauteurs\bgroup Chouzenoux\egroup\egroup{} et J.-C.
	\bgroup\fonteauteurs\bgroup Pesquet\egroup\egroup{} :
	\newblock Euclid in a taxicab: Sparse blind deconvolution with smoothed
	$\ell_1/\ell_2$ regularization.
	\newblock {\em IEEE Signal Process. Lett.}, 22(5)\string:\penalty500\relax
	539--543, May 2015.
	
	\bibitem{Sun_Q_2021_PREPRINT_convex_sbd}
	Q.~\bgroup\fonteauteurs\bgroup Sun\egroup\egroup{} et
	D.~\bgroup\fonteauteurs\bgroup Donoho\egroup\egroup{} :
	\newblock Convex sparse blind deconvolution.
	\newblock {\em PREPRINT}, juin 2021.
	\newblock \url{https://arxiv.org/abs/2106.07053}.
	
	\bibitem{Zheng_P_2023_j-ieee-spl_pendantss_pnrdantss}
	P.~\bgroup\fonteauteurs\bgroup Zheng\egroup\egroup{},
	E.~\bgroup\fonteauteurs\bgroup Chouzenoux\egroup\egroup{} et
	L.~\bgroup\fonteauteurs\bgroup Duval\egroup\egroup{} :
	\newblock {PENDANTSS}: {PE}nalized {N}orm-ratios {D}isentangling {A}dditive
	{N}oise, {T}rend and {S}parse {S}pikes.
	\newblock {\em IEEE Signal Process. Lett.}, 30\string:\penalty500\relax
	215--219, 2023.
	
\end{thebibliography}
\end{document}